\documentclass[12pt,preprint]{aastex}
\newcommand\beq{\begin{equation}}
\newcommand\eeq{\end{equation}}
\newcommand\HI{H\,{\sc i}}

\begin{document}

\title{Are Supershells Powered by Multiple Supernovae? Modeling
the Radio Pulsar Population Produced by OB Associations}

\author{Rosalba Perna\altaffilmark{1,}\altaffilmark{2} and Bryan M. Gaensler\altaffilmark{3}}

\medskip

\affil{1. Department of Astrophysical Sciences, Princeton
University, Princeton, NJ, 08544}
\affil{3. Harvard-Smithsonian Center for Astrophysics, 60
Garden Street, Cambridge, MA 02138}

\altaffiltext{2}{Spitzer fellow}

\begin{abstract}

Traditional searches for radio pulsars have targeted individual small
regions such as supernova remnants or globular clusters, or have covered
large contiguous regions of the sky. None of these searches has been
specifically directed towards giant supershells, some of which are
likely to have been produced by multiple supernova (SN) explosions
from an OB association. Here we perform a Montecarlo simulation of the
pulsar population associated with supershells powered by multiple SNe.
We predict that several tens of radio pulsars could be detected with
current instruments associated with the largest Galactic supershells (with
kinetic energies $\ga 10^{53}$ ergs), and a few pulsars with the smaller
ones. We test these predictions for some of the supershells which lie
in regions covered by past pulsar surveys.  For the smaller supershells,
our results are consistent with the few detected pulsars per bubble. For
the giant supershell GSH~242--03+37, we find the multiple SN hypothesis
inconsistent with current data at the $\sim 95\%$ level.  We stress
the importance of undertaking deep pulsar surveys in correlation with
supershells. Failure to detect any pulsar enhancement in the largest
of them would put serious constraints on the multiple SN origin for
them. Conversely, the discovery of the pulsar population associated with
a supershell would allow a different/independent approach to the study
of pulsar properties.

\end{abstract}

\section{Introduction}

Surveys for radio pulsars form the foundation for the considerable
contribution which studies of neutron stars (NS) have made to
astrophysics. Not only have such surveys identified a variety of unique
or unusual pulsars, but the large samples of objects thus accumulated
have provided vital information on the velocity distribution, luminosity
function, beaming fraction and magnetic field evolution of rotation
powered NS (e.g.\ Ostriker \& Gunn 1969; Narayan \& Ostriker 1990;
Lorimer et al. 1993).

Surveys for pulsars have usually either targeted specific regions such
as supernova remnants, globular clusters, unidentified gamma-ray sources or
steep-spectrum radio sources (e.g.\ Biggs \& Lyne 1996;
Lorimer, Lyne \& Camilo 1998), or have carried out unbiased, but usually
shallower, surveys over large regions of the sky (e.g.\ Manchester et
al.\ 1996, 2001). The former set of surveys are aimed at discovering
particulary interesting pulsars but cover only a small fraction of
the sky; the latter usually find large number of pulsars, and allow
statistical studies of pulsar birth parameters to be performed.  However,
these studies generally require apriori assumptions regarding the pulsar
birth rate and the initial spatial distribution of pulsars in the Galaxy.

In this paper, we perform the first study (to the best of our
knowledge) of pulsars correlated to supershells in our Galaxy.
Expanding giant \HI\ supershells (see e.g. Tenorio-Tagle \&
Bodenheimer 1988 for a review) have been observed for several decades
in 21 cm surveys of spiral galaxies.  These nearly spherical
structures have very low density in their interiors and high \HI\
density at their boundaries, and expand at velocities of several tens
of ${\rm km~{s}^{-1}}$.  Their radii are much larger than those of
ordinary supernova remnants, often exceeding $\sim 1$~kpc; their ages
are typically in the range of $10^6$ to a few $\times 10^7$ years.
The Milky Way contains several tens of them (Heiles 1979, 1984;
McClure-Griffiths et al. 2002), and in one case the estimated kinetic
energy of expansion is as high as $\sim 10^{54}$ ergs (Heiles 1979).
Similar supershells are also observed in other nearby galaxies (Rand
\& van der Hulst 1993; Kim et al.\ 1999).

Whereas it is clear that these \HI\ supershells result from deposition
of an enormous amount of energy into the interstellar medium, the energy
source is still a subject of debate. Collisions with high-velocity
clouds (Tenorio-Tagle 1981) could account for those cases where only
one hemisphere is present, and where the required input energy is not too
large.  However, it is unclear how such collisions could produce the
near-complete ringlike appearance observed in some cases (Rand \& van
der Hulst 1993). Loeb \& Perna (1998) and Efremov et al. (1998) suggested
that Gamma-Ray Bursts might be responsible for powering the largest supershells.
Whereas it is likely that a fraction of the smaller Galactic shells is indeed powered
by GRBs given estimates of their rates (e.g. Schmidt 1999;
Perna, Sari \& Frail 2003), however GRBs might not be able to power
the largest supershells if they are beamed (and therefore their energies reduced
compared to their isotropic values $\sim 10^{53}-10^{54}$ ergs). 

The most discussed interpretation in the literature is that
supershells are a consequence of the collective action of stellar
winds and supernova (SN) explosions originating from OB star associations
(e.g. McCray \& Kafatos 1987; Shull \& Saken 1995).  Several hundreds
of stars exploding within a relatively short time are necessary to power
the most energetic supershells.

The motivation for simulating and studying the population of pulsars
associated with supershells is therefore twofold: first, it would
provide an independent test of the multiple SN origin for them;
second, if the pulsar population associated with a supershell is
indeed identified, it would possibly provide an independent way to
constrain pulsar birth parameters. This is because both the birth rate
and the birth location of the pulsars would be independently
constrained rather than assumed a priori.

The paper is organized as follows: \S 2.1 discusses the model for the
superbubble growth and the corresponding pulsar birth rates; the
distributions of the pulsar characteristics and the Montecarlo
simulation of the observable population are discussed in \S 2.2. In \S
3, the results of our simulations are compared to the data for the
supershells which lie in regions of sky covered by past, sensitive
pulsar surveys, and predictions are made for some of the other supershells for which
current data are not sufficient to make a comparison at this time. 
Our work is finally summarized in \S 4.

\section{Simulating a population of pulsars from an OB association}

\subsection{OB associations and supershell growth}

\setcounter{footnote}{3}

The evolution of a superbubble powered by a continuous energy source\footnote{
The approximation made here is that each SN, rather than outputing its energy
suddenly at time $t$, spreads it output over an interval $\Delta t$ until
the next explosion. Mac Low \& McCray (1988) showed that after five to ten
SNe
have exploded, the continuous energy injection model provides a very
good approximation to the discrete one.}
has been calculated by Weaver et al. (1977; see also MacLow \& McCray 1988;
Shull \& Saken 1995). Weaver et al. derived a similarity solution in terms
of the equivalent mechanical luminosity of supernovae, 
$L_{\rm SN}$, ambient density $\rho_0$ and time $t$. This yields
the time-evolution of the superbubble radius
\beq
R = \left(\frac{125}{154\pi}\right)^{1/5}L_{\rm SN}^{1/5}\rho_0^{-1/5}t^{3/5}
\approx (267 {\rm pc}) \left(\frac{L_{38}t_7^3}{n_0}\right)^{1/5}\;,
\label{eq:rad}
\eeq
and velocity
\beq
V\approx (15.7 {\rm km s^{-1}}) L_{38}^{1/5}n_0^{-1/5}t_7^{-2/5}\;.
\label{eq:vel}
\eeq
Here $n_0$ is the atomic number density (in cm$^{-3}$),
$t_7=t/(10^7{\rm yr})$ and $L_{38}=
L_{\rm SN}/(10^{38}{\rm ergs~s^{-1}})$ is the equivalent of one SN of
energy $E_{\rm SN}=10^{51}$ ergs occurring every $\Delta t_{\rm SN}=E_{\rm SN}/L_{38}=
3.2\times 10^5$ yr.\footnote{Note that the numerical value of the 
interval $\Delta t_{\rm SN}$ has no special meaning but it is only resulting from
the definition of $L_{\rm SN}$ in units of $L_{38}$.}
For a given superbubble, from the observed values of $R$, $v$ and $n_0$,
we derive the corresponding age $t_7$ and the equivalent mechanical luminosity
$L_{\rm SN}$. The number $N_{\rm SN}$ of stars that become supernovae (SN) over
the time scale $t$ is given by $N_{\rm SN}=L_{\rm SN}t/E_{\rm SN}$. 
Although stellar winds from the OB association will produce a hot bubble
before the first SN goes off, these winds are not important compared to SNe for the
later dynamics of the supershell (McCray \& Kafatos 1987; Mac Low \& McCray 1988). 
Also, while the early times of the bubble evolution are sensitive
to the details of the rate of energy injection (which is assumed 
constant in the Weaver et al. model), the late phase of the bubble
is not\footnote{There might however be discrepancies if a considerable
SN activity occurs near the shell rim (Oey 1996).} (Shull \& Saken 1995). 
We should point out that the factor $N_{\rm SN}$ estimated using the equations
above is typically a factor of {\rm a few smaller} than the equivalent
number estimated using the energy $E_{\rm E}=5.3\times 10^{43}
n_0^{1.12}R^{3.12}V^{1.4}$~ergs which would be required
to produce the shell by a sudden explosion (Chevalier 1974). The latter,
larger, number has been the one often used in the literature to estimate
superbubble energies (Heiles 1979; Rhode et al. 1999; McClure-Griffith et al. 2002).

The initial mass function for massive stars can be written as 
(Garmany, Conti \& Chiosi 1982)
\beq 
f_{\rm IMF}(M_*)\equiv dN_*/dM_*\propto M_*^{-\beta}\;,  
\label{eq:IMF}
\eeq
where $\beta\sim 2.6$. We assume $M_{\rm min}=3 M_\odot$ and
$M_{\rm max}=140 M_\odot$ for the minimum and maximum mass of the distribution, 
respectively.

The main-sequence lifetimes of massive stars are given
approximately by (Stothers 1972; Chiosi, Nasi \& Sreenivasan 1978)
\beq
t_*\sim \left\{
  \begin{array}{ll}
3\times 10^7(M_*/10M_\odot)^{-1.6}\; {\rm yr} \;\;
& \hbox{if $\;7\;\la M_*\la 30 M_\odot$}\\   
 9\times 10^6(M_*/10M_\odot)^{-0.5} \;{\rm yr} \;\;
& \hbox{if $\;\;M_*\ga 30 M_\odot$} \\
\end{array}\right.\;.
\label{eq:tMS}
\eeq
The least massive star that is expected to terminate as a Type II SN
has initial mass $M_{\rm min, SN}=7 M_\odot$ (Trimble 1982). We take
$M_{\rm max}\sim 140 M_\odot$ as the mass of the most massive star of
the association. Our results are rather insensitive to the precise value
of the minimum star mass, $M_{\rm min}$, in the distribution, as stars
with $M<M_{\rm min, SN}$ would not contribute to the energetics of the shell.
Similarly, our results are also rather insensitive to the precise
value of the upper mass cutoff due to both the steep decline in $dN_*/dM_*$
at large $M_*$, as well as to the fact that low metallicity stars with
mass in the range $40-140 M_\odot$ collapse directly to a black hole (BH)
without any significant explosion energy\footnote{A fraction of these
stars with a sufficient amount of angular momentum might explode as
hypernovae and give rise to gamma-ray bursts. In those particular, but rare
cases, an amount of energy would be released into the medium, but
probably not significantly larger than that of a standard SN.} (Heger
et al. 2003). 
The maximum mass of the star that will leave behind a
NS is rather uncertain. However, estimates of rates of heavy element
production require stars as massive as 25 $M_\odot$ to form supernovae
(e.g., Hillebrandt 1982).  Numerical simulations (e.g., Heger et
al.\ 2003) find that for low-metallicity, non-rotating isolated stars, $M=25
M_\odot$ is the maximum initial star mass that would leave behind a
NS star remnant, while stars in the range $25-40 M_\odot$ form a
BH by fallback while releasing a small amount of energy into
the medium. Following these results, here we assume that only stars in
the range $7-25 M_\odot$ leave behind a NS, while those with
$M>M_{\rm SN,max}= 25M_\odot$ (but $M< 40 M_\odot$) contribute to
power the supershell, but do not leave behind a pulsar. For stars with
mass below 7 $M_\odot$ we assume that their energy contribution is
negligible compared to that of the SNe. Note that, if stars had solar
metallicity, they would be expected to form NS remnants for a much larger
range of initial masses (Heger et al. 2003), and therefore there would be many more
pulsars than what we predict here\footnote{As a matter of fact, there is a large spread in the
metallicity of stars in the Galaxy (e.g. McWilliam 1997),
from much lower than solar to about solar. The assumption 
of low metallicity for all the stars of the association is a conservative 
assumption for our purposes.}.  Finally, we consider a model of an OB
association with coeval star formation; that is all stars are assumed
to be formed at once with no age spread. Once again, note that this is
a conservative assumption for our predictions, because if stars were
formed with an age spread, the pulsars in the bubble would be
generally younger than in the coeval case, and hence would be more
likely to be detectable (as younger pulsars have typically larger luminosities and
beaming fractions). Also, they would have had less time to travel far
from their birthplaces, and therefore a larger fraction of them
would be closer to the supershell at the present time. 

\subsection{Properties of the pulsar population}

Modelling the intrinsic properties of the Galactic population of
pulsars, especially with respect to the luminosity function and
initial spin period, has been the subject of extensive investigation
in the past few decades (e.g. Gunn \& Ostriker 1970; Vivekanand \&
Narayan 1981; Phinney \& Blandford 1981; Narayan \& Ostriker 1990;
Lyne et al. 1985; Narayan 1987; Stollman 1987; Emmering \& Chevalier
1989; Lorimer et al. 1993; Johnston 1994; Cordes \& Chernoff 1998).
Despite all these efforts, no consensus has been reached on what the
intrinsic and birth properties of the pulsars are. The most recent,
comprehensive analysis, based on large-scale 0.4 GHz pulsar surveys,
has been made by Arzoumanian, Cordes \& Chernoff (2002; ACC in the
following).  Here, as a first step in estimating the expected number
of pulsars for a supershell of given age and energy, we use
their inferred parameters under the assumption that spin-down is only
caused by dipole radiation losses.  They find that pulsars are likely
to be born with velocities distributed according to a two-component gaussian
distribution, with $40\%$ of them having a characteristic speed
(i.e. magnitude of the 3-D velocity)
of 90~km~s$^{-1}$ and the remaining  of 500~km~s$^{-1}$.  The initial birth
period distribution (taken as a gaussian in log), has a mean $\langle \log
P_0(s) \rangle =-2.3$ with a spread greater than $0.2$\footnote{Here we adopt the value
0.3, after verifying that variations in the range 0.3-0.6
do not give statistically different results.}, while the initial magnetic field
strength (also a log-gaussian), is found to have a mean $\langle \log
B_0[G] \rangle =12.35$ and a variance of $0.4$.  These distributions of
initial parameters have been derived under the assumption of no
magnetic field decay. The spin evolution of the pulsars is then simply
given by:
\beq 
P(t) = \left[P_0^2 + \left(\frac{16\pi^2
R^6 B^2}{3Ic^3}\right)t\right]^{1/2}\;,
\label{eq:spin} 
\eeq
where $I=10^{45}$~g~cm$^2$ is the moment of inertia of the star, 
and $R=10$~km its radius.

For simplicity and because the ACC radio luminosity model is not obviously
scalable to frequencies other than 400 MHz, 
here we adopt the luminosity function proposed by Emmering \& Chevalier (1989)
\beq
L_\nu= \gamma P^{-1.61}\dot{P}^{0.5}\;.
\label{eq:lum}
\eeq
Lorimer et al. (1993)  found that the best fit to the data at $\nu = 400$~MHz was obtained
with $\gamma=3.5 $ mJy kpc$^2$, which is the value we adopt here. 
The flux density for a pulsar at a distance $D$ is related to its luminosity
through the relation $S_\nu=L_\nu D^2$. 
We determine the coefficient
of the luminosity function at other frequencies assuming a typical spectral
index $\alpha=- 1.8$, where $L_\nu \propto \nu^\alpha$ (Maron et al.\ 2000). 
For a given $P$ and $\dot{P}$, we also allow for a spread in luminosity
by considering a log-gaussian probability distribution with mean $\langle L
\rangle$ given by Equation~(\ref{eq:lum}) 
and spread $\sigma_L$. As pointed out by ACC,
this is a way to parameterize the fact that the pulsar
luminosity has an angle dependence with respect to the magnetic axis,
and the particular value that is observed by a given observer
depends on its orientation with respect to that axis. 
We find that, with a mean given by Equation~(\ref{eq:lum}),
a spread $\sim 0.8$ (in the log) appears to give results consistent with the
current data.  
  
For the evolution of the beaming fraction $f$ with period we adopt
the relation derived by Tauris \& Manchester (1998)
\beq
f=0.09 \left(\log\frac{P}{10}\right)^2 +0.03\ ,
\label{eq:fbeam}
\eeq
where $P$ is in seconds.

Pulsars are radio active until they slow down to a point where 
the electric field near the polar cap becomes too small (e.g. Chen \&
Ruderman 1993). The transition for a given $P$ and $\dot{P}$ is
probably not abrupt, and rather than being described by a ``death line'',
it is probably better described by a ``death band''. Following
ACC, we define the probability that objects with
given $\dot{P}/P^3$ are radio-active as
\beq
\frac{1}{2}
\left\{ 
\tanh\left[\left(\dot{P}^3_{15}/P^3-10^{\rm
DL}\right)/\sigma_{\rm DL}\right] +1 
\right\}\;,
\label{eq:active}
\eeq
where $\dot{P}_{15}\equiv \dot{P}/(10^{-15}{\rm
s}~{\rm s}^{-1})$. DL$= 0.5$ and $\sigma_{\rm DL}= 1.4$ represent the position
and lateral extent of the death band, respectively.  

For the pulsar velocity distribution at birth, we adopt the
double-gaussian model of ACC described above. The orientation of the
pulsar kick at birth is assumed to be random, and the Galactic
rotational velocity (220~km~s$^{-1}$) is added vectorially to the
initial kick velocity.

Finally, we compute the detector sensitivity using the standard detector
equation (e.g.\ Dewey et al.\ 1984):
\beq
S_{\rm lim} = \frac{\sigma\beta (T_{\rm sys} + T_{\rm sky})}{G\sqrt{B N_p\tau_{\rm
obs}}}\,\left(\frac{W}{P-W}\right)^{1/2}\;,
\label{eq:det}
\eeq
In the above equation, $\sigma$ is a loss factor, 
$\beta$ is the signal-to-noise ratio for a detection
(assumed to be 8), $T_{\rm sys}$ is the system temperature, $G$ the
telescope gain, $B$ the receiver bandwidth, $N_p$ the number of polarizations
and $\tau_{\rm obs}$ the observation time. The specific values of these
parameters for the various surveys we consider
can be found in the papers referenced in \S3 below.
The observed pulse width is given by
$W=(W^2_{50}+\tau_{\rm samp}^2 + \tau_{\rm DM}^2 + \tau_{\rm
scatt}^2)^{1/2}$,
$W_{50}$ is the intrinsic pulse width, assumed equal to 5\% of the period,
$\tau_{\rm samp}$ is the sampling interval, $\tau_{\rm DM}$ is the
dispersion smearing across one frequency channel, and $\tau_{\rm scatt}$
is the broadening of the pulse due to interstellar scattering.
The dispersion measure and scattering measure of each pulsar 
(needed to calculate the above quantities) are
computed using the public code\footnote{http://astrosun.tn.cornell.edu/\~{ }cordes/NE2001/} 
by Cordes \& Lazio (2003), while the all-sky data are from the 408 MHz
compilation of Haslam et al (1982), scaled to other frequencies 
assuming a spectral index of --2.6 for the brightness temperature 
of the diffuse Galactic synchrotron background.
We neglect the effects of cut-offs in harmonic summing and
filtering of the data (see e.g. Manchester et al. 2001), 
which has a small effect on the sensitivity for periods which are
not very short and for large DMs (F. Crawford 2003, private communication).  

\section{Simulating a survey of pulsars in a supershell}

The simulation of the pulsar population produced by the OB association
that gave rise to a supershell is of the Montecarlo-type.  First,
the mass of a star is generated from the distribution in
Equation~(\ref{eq:IMF}); only stars whose lifetime is shorter than the present age of
the supershell, $t_{\rm shell}$, are followed. The birthtime $t_{\rm birth}$ of a given
pulsar is taken as the time that the progenitor star explodes.  For
each pulsar, the birth properties ($B$ field, initial velocity, initial
period) are computed according to the ACC distribution described
above.  The period of each pulsar is then evolved for a time $t_{\rm
puls}=t_{\rm shell}- t_{\rm birth}$ according to Equation~(\ref{eq:spin});
if the pulsar is still above the death band, a luminosity is assigned
(based on the probability distribution described in \S 3). 
To determine whether the pulsar is beamed towards our line of sight, a
random number $x$ in the interval \{0,1\} is generated. If $x\le f$
(defined in Eq:(\ref{eq:fbeam}), then the pulsar is considered beamed towards us.
The birth place of the pulsars is assumed to coincide with the center of the
supershell; however, because of the relative motion of the supershell
with respect to the sun (due to Galactic differential rotation), the
coordinates of the supershell on the sky are a function of time.  We
account for this effect for a pulsar of age $t_{\rm
puls}$ by extrapolating backwards by a time $t_{\rm puls}$ to determine
the coordinates of the supershell on the sky (and hence of the pulsar
birth place) at that specific time in the past.  The final position of the
pulsar (i.e., at its current age) is determined by integrating its
equations of motion in the Galactic potential (Kuijken \& Gilmore 1989)
for a time $t_{\rm puls}$.
Finally, the flux of each pulsar is compared to the
detector threshold after computing the dispersion and scattering
measure for its position.  The simulation is stopped when the
number of supernovae reaches what is needed to power the supershell.

Galactic shells and supershells have been studied for several decades
since the pioneering work of Heiles (1979, 1984).  Pulsar surveys, on
the other hand, have never been specifically directed at sampling
supershell regions at high sensitivity. Therefore, for the purpose of
our analysis in this paper, we use the results of existing surveys,
while emphasizing the importance of future, deeper surveys of the
supershell regions.

A Montecarlo realization of the 1.4-GHz pulsar population expected to
be associated with the two Galactic supershells GSH~242--03+37 and
GSH~088+02--103 (which are among the biggest in the Galaxy) is shown
in Figure 1 and 2.  In both cases, the left panel shows the expected
luminosity versus age distribution of pulsars beamed towards
us (independently of whether still inside the shell or not).  The
oldest pulsars have an age comparable to the age of the
supershell. The right panel shows the current position
on the sky of the pulsars with respect to that of the supershell.
Note that some of the pulsars have moved very far from the supershell
and therefore they do not appear in the right panels of the figures
(which shows the region in the sky around the supershell).  Our
simulations show that there is a significant probability of finding an
enhancement of pulsars in the innermost regions of the bubble. These
pulsars are a combination of the youngest sources in the sample, those
on the low-velocity tail of the distribution, and those whose velocity
vector happens to be almost parallel to the line of sight.  The region
where GSH~088+02--103 lies has been searched in the Green Bank survey
at 370 MHz (Sayer, Nice \& Taylor 1997), while GSH~242--03+37 has been
searched at 436 MHz by the Parkes Southern Pulsar survey (Manchester
et al.\ 1996), and also had a few pointings at 1.4 GHz in the Parkes
Multibeam Pulsar survey (Manchester et al. 2001; A.\ J.\ Faulkner,
2003, private communication). The predicted number of detected pulsars
for the survey parameters corresponding to the region around each
supershell are listed in Table 1. In Figures 1 and 2, as an example,
we show the pulsars which would be visible with survey parameters
corresponding to those of the current Parkes Multibeam Pulsar survey
(filled symbols), as well as those which could not be detected by
Parkes but would be visible to the planned Square Kilometer Array
(SKA), which will have a sensitivity of about~1.4 $\mu$Jy for one
minute of integration time (Kramer 2003).

For a quantitative comparison with current data, we selected the most
energetic supershells among those at low Galactic latitude (Heiles
1979; McClure-Griffiths et al. 2002), where most of the pulsar surveys
have been focused.  Our results are summarized in Table 1.  For each
supershell, we ran 1000 Montecarlo realizations of the pulsar
population associated with it, and determined the mean and variance of
the distribution for the number of pulsars that are expected to be
found within the supershell region on the sky at the current time. To
be more specific, if $l_0,b_0$ are the Galactic coordinates of the
supershell, and $\Delta l$ and $\Delta b$ its total extent
(i.e. diameter) on the sky, then the pulsars associated with the
supershells are those whose coordinates $l,b$ satisfy the condition
$|l-l_0|<\Delta l/2$ and $|b-b_0|<\Delta b/2$, and are within a distance
$D\pm R$, where $D$ is the distance to the shell and $R$ its mean radius.
These are only a
fraction of the ones produced within the association and potentially
observable, as shown in Fig. 1 and 2. For each supershell, we computed
the distribution of pulsars whose flux is above the threshold of the
deepest survey which has covered that particular region of sky in
which the shell lies, as well as the distribution of pulsars which
would be detectable with SKA. The pulsars in the current
catalogue\footnote{http://www.atnf.csiro.au/research/pulsar/psrcat/}
which are candidate associations are also reported. These are selected
on the basis of their coordinates $l,b$ which have to satisfy the
condition $|l-l_0|<\Delta l/2$ and $|b-b_0|<\Delta b/2$, as well as on
their distance, which has to be within $D\pm R$. However, both the pulsars
and the shell distances are known with some uncertainty. For pulsars,
the distance as inferred from dispersion measure has some 25\%
uncertainty on a statistical average\footnote{In some cases distances
estimated from the DM may be off by factors of $\sim 2$ or greater;
however, given that ours is a statistical analysis, we adopt the
statistical average uncertainty for the model as a whole.} 
(see e.g. Cordes \& Lazio 2003), while the errors on the
supershell distances are specifically indicated in the
McClure-Griffiths et al. sample, and most of them are in the range
10-20 \%. Unfortunately no distance errors are provided in the Heiles
(1979) sample, and for those we assume a more conservative 25\% uncertainty.  We then
consider a pulsar a possible association with the supershell if it
falls within $D\pm R$ after accounting for the distance uncertainties
in both the pulsar and the supershell. 

As Table~1 shows, in a few cases, due to a combination of the large
distances of the supershells and the comparatively poor sensitivity
level at which the corresponding region was surveyed, no pulsars are
expected to have been detected in association with the supershell, and
therefore no meaningful constraints can be derived.  For the two
supershells which lie in the region covered by the Parkes Multibeam
survey (GSH~285--02+86 and GSH~292--01+55; McClure-Griffiths et al. 2002),
the predictions of the multiple SN scenario at the detection level of
that survey are consistent with the number of candidate associations,
although they do not rule out other interpretations, such as GRBs or
collisions with high-velocity clouds, as in both these scenarios one
would not expect any pulsar enhancement correlated with the
supershell.  Deeper surveys of those regions are needed to test
the multiple SN scenario.
However, an interesting constraint can already be derived for
GSH~242--03+37 (Heiles 1979). Our results (see Table 1) show that there is
a 95\% probability that the supershell was not the result
of multiple SNe. Deeper surveys are necessary to set much tighter
constraints. Our current results, in fact, are rather dependent
on the luminosity function (Lorimer et al. 1993) that we have adopted here;
however, recent new discoveries of low-luminosity radio pulsars
(Camilo et al. 2002a, 2002b) might indicate that young pulsars
might be fainter than previously realized, and this could be an
alternate explanation for the lack of pulsars in the supershell at
the current sensitivity threshold. However, when the sensitivity
is sufficiently high that even the faintest pulsars can be detected 
(as it will be with the planned SKA instrument), then the 
particular details of the luminosity function will be irrelevant for 
the proposed experiment.      
Also note that, while current pulsar surveys are able to
probe mostly the pulsars within our Galaxy, the SKA survey
will be able to detect a large fraction of pulsars also in the LMC, which
has several giant supershells (Kim et al. 1999). The analysis that
we are proposing here can therefore be made also for the
LMC supershells  in the SKA era.

Before concluding we should note that, besides depending on the total
energetics\footnote{We recall that the Weaver et al. (1977) solution that
we have adopted yields an energy  value that is a few times smaller
than the Chevalier (1974) solution for sudden injection, often used
in the literature for supershells. Adopting the latter
model would entail us predicting many more supernova explosions,
and hence detected radio pulsars, than calculated above.}
of the supershells (which determine the total number of SN
explosions needed), our results also rely on the pulsar birth
parameters, and in particular on their velocity distribution, magnetic
field, and periods. Here we have adopted the ACC results for
the case with braking index $n=3$, which corresponds to pure dipole
losses. In reality, braking indices are often found to be different
than 3 (e.g. Lyne et al. 1996), possibly due to a variety of causes 
(e.g. Menou et al. 2001 and references therein), and this
affects the pulsar evolution. ACC determined the initial birth
parameters also for the case $n=4.5$, and found that they were not
significantly different. In particular, the velocity distributions for
the two cases are consistent within the error bars. Given that both
sets of initial parameters are determined so that they yield the same
pulsar distribution today, different initial conditions should not
affect much our predictions for the observability of pulsars in the
bubbles today. At any event, future population studies resulting
from the analysis of the Parkes Multibeam Survey will be very important
for further constraining the underlying pulsar distributions (Lorimer 2003; 
Vranesenic et al. 2003) and,  
as a matter of fact, the study of the pulsar population
associated with the largest supershells can be potentially used as an
independent way to constrain pulsar birth parameters.

Finally,  we note that, if the giant supershells are indeed powered by
multiple SN explosions, then besides the several tens of NS visible as
radio pulsars, there should be hundreds of other NSs (all the pulsars
not beamed towards us and all the ones that have passed the death line), plus a
smaller number of BHs, in the region delimited by the superbubble. The
only hope to possibly detect such large concentrations of compact objects
would be through accretion from the interstellar medium on a Bondi-type
mode (see Blaes \& Madau 1993 for NSs and Agol \& Kamionkowski 2002 for
BHs). However, current limits for detection of isolated, accreting NSs,
may suggest that accretion occurs at a rate which is likely well below
the Bondi value (Perna et al. 2003; Toropina et al. 2003), and therefore
this large concentration of isolated NSs and BHs in supershells might
have to await more sensitive X-ray instruments to be possibly detected
in X-rays.

\section{Summary}

Pulsar searches so far have mostly focused either on 
specifc regions such as supernova remnants and globular clusters or on
a large contiguous region of sky.  In this paper we have, for
the first time, simulated the pulsar population that should be
associated with the giant supershells observed in our Galaxy and in
nearby galaxies, if their interpretation as due to multiple SN
explosions is correct.  We have found that, for the largest
supershells, several tens of pulsars should be present within the
supershell region.

Failure to detect any pulsar enhancement in the largest superbubbles would
put serious constraints on the multiple SN origin for them. 
Conversely, the discovery of the pulsar population associated with
a superbubble would provide new insights into the
distribution and birth properties of pulsars. 

\acknowledgements We thank Andy Faulkner and Froney Crawford for information on the Parkes
Multibeam Pulsar Survey, and Mordecai-Mark MacLow for a very useful
discussion on estimates of supershell energetics.
We also thank the referee, Fernando Camilo, for a careful reading of our
manuscript and helpful comments.

\newpage

\begin{deluxetable}{lcccccccccc}

\tablewidth{0pt}
\tablecaption{Selected Galactic supershells and possibly associated pulsars
according to the criteria described in the text.}
\tablehead{\colhead{Name} & \colhead{$D$\tablenotemark{(a)}}&\colhead{$\Delta l$
\tablenotemark{(b)}}&\colhead{$\Delta b$ \tablenotemark{(c)}}
&\colhead{$t_{\rm shell}$ \tablenotemark{(d)}}&\colhead{$N_{\rm SN}$
\tablenotemark{(e)}}&
\colhead{$\langle N_{\rm cur}\rangle$\tablenotemark{(f)}}& 
\colhead{$\sigma_{N,{\rm cur}}$} & 
\colhead{$\langle N_{\rm SKA}\rangle $\tablenotemark{(g)}}& 
\colhead{$\sigma_{N,{\rm SKA}}$} 
&\colhead{$N_{\rm puls}$ 
\tablenotemark{(h)}} }
\startdata		    
GSH~022+01+139  & 9.5 &4 &3 & 1.0&43 & 0.43 &0.63& 1.4&1.1& 2  \\
GSH~064--01--97  & 16.9&11 &5&3.3& 403 &0.016&0.12&8.3&2.8 &0 \\
GSH~088+02--103 &12.6&7 &5 &1.54& 480 &0.001&0.032&17.0&3.9 &0 \\
GSH~095+04--113 &12.9& 10 &5 &4.6 &84 & 0& 0& 1 &1 & 0 \\
GSH~242--03+37 &3.6& 15 &15 &1.4 & 664 &7.3 &2.8 &21.4 &4.5 & 2 \\
GSH~285--02+86  &13.7 & 3.3 &3.2 & 1.1 &90 &0.86 &0.89& 3.4& 1.7& 2  \\
GSH~292--01+55  & 11.6 & 5.1 &2.0 &1.44 &200 &1.5 &1.3&4.9 &2.1& 2  \\ 
\enddata
\tablenotetext{a}{Distance to the supershell in kpc. The values are taken
from McClure-Griffith et al. (2002) for GSH~285--02+86 and GSH~292--01+55 
and from Heiles (1979) for the other ones.}
\tablenotetext{b} {Angular extent of the supershell in longitude.}
\tablenotetext{c} {Angular extent of the supershell in latitude.}
\tablenotetext{d} {Age of the supershell in units of $10^7$ yr.}
\tablenotetext{e}{Number of SN explosions required to power the supershell.}
\tablenotetext{f}{Mean number of pulsars predicted within the supershell volume
and above the sensitivity
of the deepest survey that has sampled the supershell region; $\sigma_{N,{\rm cur}}$
is the standard deviation of that distribution. The surveys for which
the simulations have been made are: the Parkes Multibeam Pulsar 
survey for GSH~022+01+139, GSH~285--02+86, and GSH~292--01+55;  
the Green Bank Northern sky survey for GSH~088+02--103, and GSH~095+04--113 and
the Parkes Southern Pulsar survey for GSH~242--03+137. GSH~064--01--97 has been
partially covered by Arecibo (Nice et al. 1995).}
\tablenotetext{g}{Mean number of pulsars predicted within the supershell
volume above the SKA sensitivity for one minute of integration time (Kramer 2003);
$\sigma_{N,{\rm SKA}}$ is the standard deviation of the distribution.} 
\tablenotetext{h}{Number of candidate pulsar-supershell associations in the
current pulsar catalogue.}

\end{deluxetable}

\clearpage
\newpage

\begin{figure}[t]
\plottwo{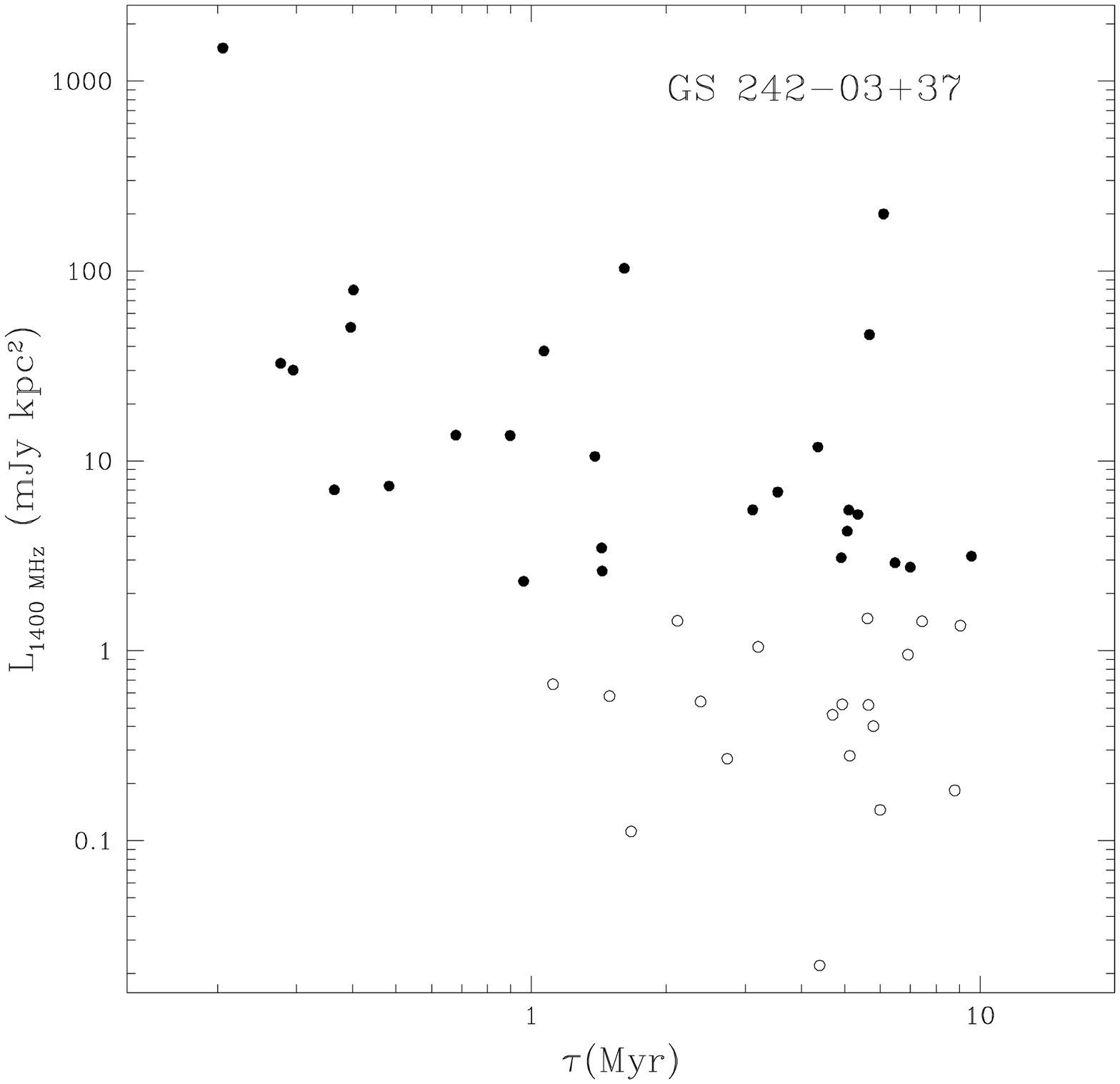}{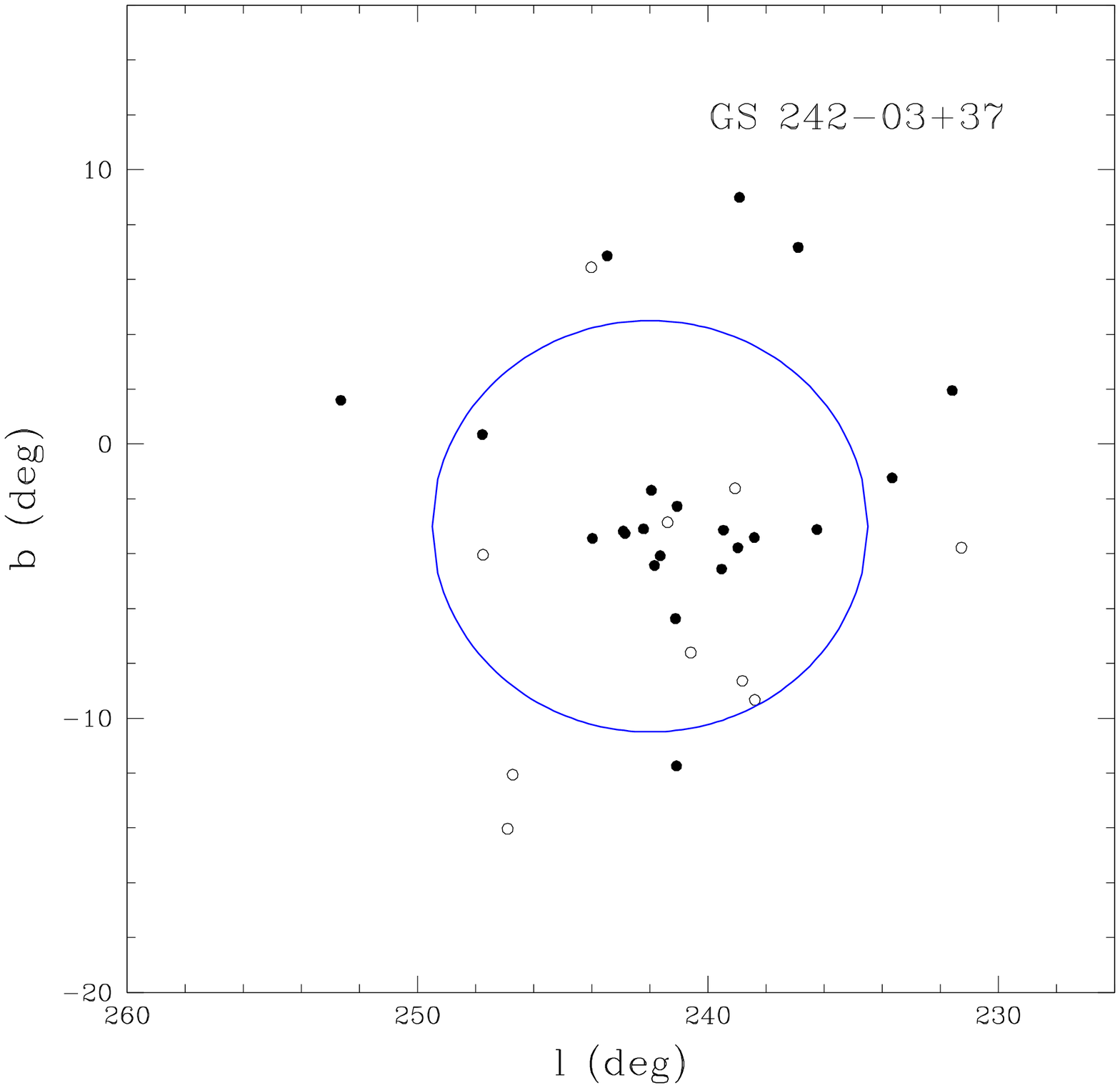}
\caption{The detectable pulsar population predicted to be
associated with the Galactic
supershell GSH~242--03+37. Filled circles represent pulsars above the
sensitivity limit corresponding
to the parameters of the 1.4-GHz Parkes Multibeam Pulsar survey, 
while open circles correspond to pulsars which could
be detected in a 1.4-GHz pulsar survey using the Square Kilometer
Array.  The left panel shows the distribution of pulsar luminosities and ages
for all pulsars formed in the supershell which are above the death line
and beamed towards us, while 
the right panel shows their projected position on the sky 
with respect to the position of the supershell (whose approximate
boundary is indicated by the ellipse).}
\end{figure}

\begin{figure}[t]
\plottwo{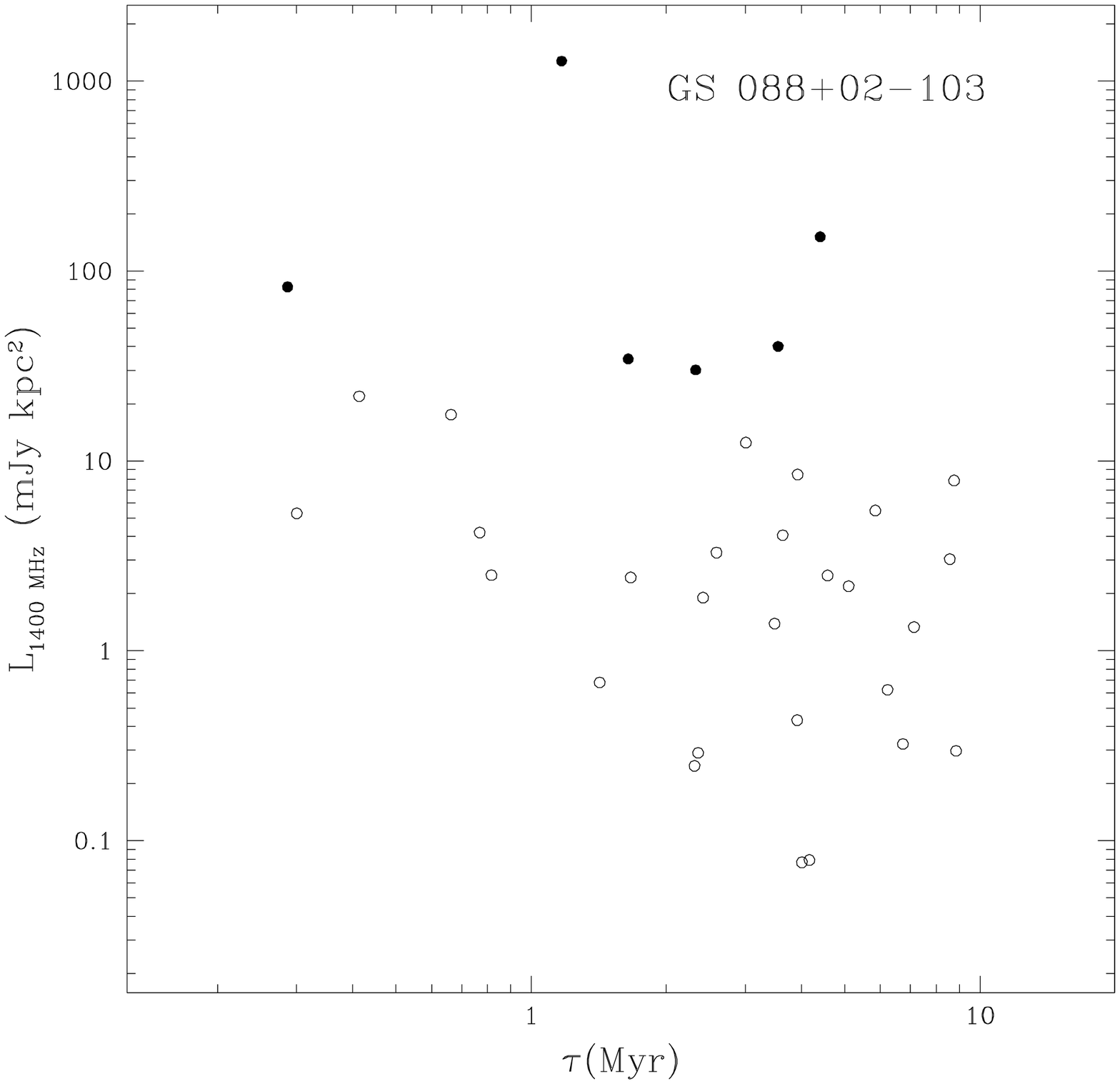}{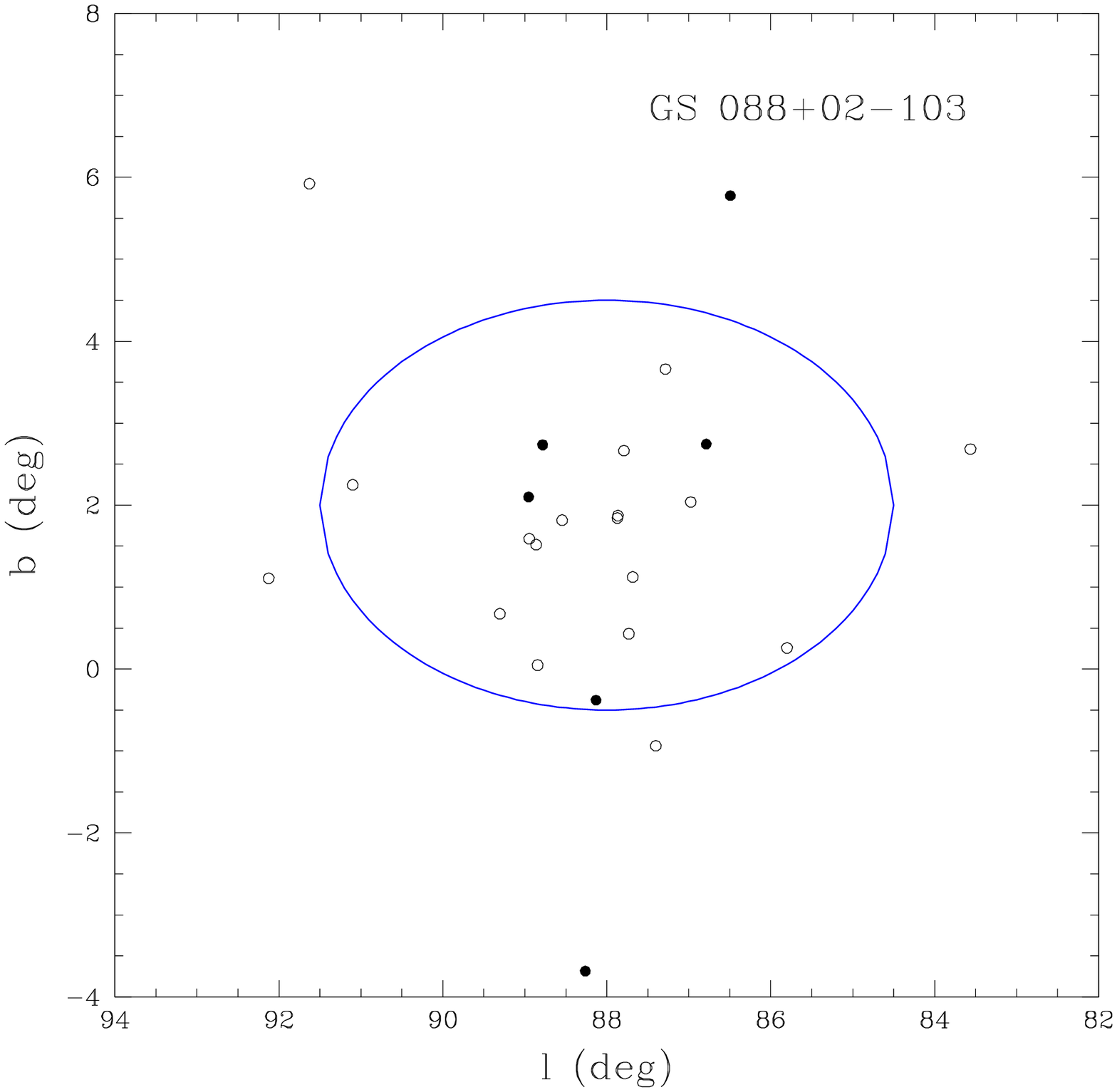}
\caption{Same as for Fig.\ 1, but for the Galactic supershell GSH~088+02--103.}
\end{figure}

\end{document}